\documentclass[aps, showpacs, pra, superscriptadaddress, twocolumn, amssymb 
] {revtex4-1}
\usepackage{hyperref}
\usepackage{amsmath }
\usepackage{amssymb}
\usepackage{epsfig}
\usepackage{natbib}
\usepackage{epstopdf}
\usepackage{graphicx}

\usepackage{floatrow}
\usepackage[caption=false,font=Large]{subfig}

\newcommand*{\be}{\begin{equation}}
\newcommand*{\ee}{\end{equation}}
\newcommand*{\bea}{\begin{eqnarray}}
\newcommand*{\eea}{\end{eqnarray}}

\newcommand*{\R}{\mathcal{R}}

 \DeclareFontFamily{OT1}{pzc}{}
 \DeclareFontShape{OT1}{pzc}{m}{it}%
 {<->  s  *  [1.400]  pzcmi7t}{}
\DeclareMathAlphabet{\mathscr}{OT1}{pzc}%
{m}{it}

\begin{document}

\title{
Hamiltonian formulation of the standard $\mathcal{PT}$-symmetric nonlinear Schr\"odinger dimer}

\author{I V Barashenkov}
  \affiliation{Centre for Theoretical  and Mathematical Physics,  University of Cape Town,    Rondebosch 7701, South Africa  }

\begin{abstract}
The standard  $\mathcal{PT}$-symmetric dimer is  a linearly-coupled
two-site discrete nonlinear Schr\"odinger equation 
with one site losing and the other one gaining energy at the same rate. 
We show that despite gain and loss, the standard $\mathcal{PT}$-dimer  is a Hamiltonian system.
 We also produce  a 
  Lagrangian formulation for the dimer.
\end{abstract}

\maketitle

\section{Introduction} 

The  Schr\"odinger dimer is a two-site discrete nonlinear
Schr\"odinger equation of the form 
\begin{align*}
i \frac{du}{dz}+v    +  F(u,u^*,v,v^*)=0, \nonumber  \\
i \frac{dv}{dz} +u+ G(u,u^*, v, v^*)=0.
\end{align*}
The physically interesting situation pertains to the functions
 $F$ and $G$  being covariant under the simultaneous phase shifts in $u$ and $v$; 
  that is, 
\begin{align*}
F(e^{i \alpha}u, e^{-i \alpha}u^*,  e^{i \alpha}v,   e^{-i \alpha}  v^* )  = & e^{i \alpha} F (u,u^*,v,v^*), \\
G (e^{i \alpha}u, e^{-i \alpha}u^*,  e^{i \alpha}v,   e^{-i \alpha}  v^* )  = & e^{i \alpha} G (u,u^*,v,v^*)
 \end{align*}
for any real $\alpha$.

The Schr\"odinger dimers with various polynomial nonlinearities are workhorses of  photonics, where they serve to model stationary light beams 
 in  coupled optical waveguides \cite{couplers,Maier_Snyder}.    
 The variables
$u$ and $v$ represent the corresponding complex beam amplitudes,  
and $z$ measures the distance along the parallel cores. 

 Dimers  also occur in  the  studies of the 
Bose-Einstein condensate trapped in a double-well potential \cite{BEC,Theocharis}. 
Here,  $u$ and $v$ are the complex amplitudes of the
  mean-field condensate
wave functions localized in the left and right well, respectively
\cite{BEC},  or the
amplitudes of the ground 
and the first excited state \cite{Theocharis}.
 The nonlinear Schr\"odingier dimers were employed, extensively,
 in the solid state physics ---  where they give the simplest  discrete self-trapping equations \cite{Eilbeck,DST,Tsironis_review} ---
and in the context of electric lattices \cite{Tsironis_review}. 

With the advent of the parity-time ($\mathcal{PT}$) symmetry \cite{PT}, the studies of the optical couplers expanded to include 
structures 
 consisting of a waveguide with loss and a guide with an equal amount of optical gain.
 While a variety of cubic nonlinearities were considered, 
 the most   frequently used nonlinear model 
 has the form \cite{Maier_Snyder,PT_couplers,Kip,Dimer_integrability,SXK,lattices}
 \begin{align}
i \frac{du}{dz} + v+ |u|^2 u   & =   \phantom{-}   i \gamma u,  \nonumber  \\
i \frac{dv}{dz}  +u+ |v|^2v   &  =  -   i\gamma v.
\label{A3}
\end{align} 

This discrete nonlinear Schr\"odinger equation  
 is occasionally referred to as the {\it standard\/}  $\mathcal{PT}$-symmetric dimer. 
Here, the quantities $P_1=|u|^2$
and $P_2=|v|^2$ give
the powers carried by the active and lossy channel,
respectively, and $\gamma>0$ is the corresponding gain-loss rate.
The model also has an interpretation in the matter-wave context 
where it emulates a $\mathcal{PT}$-symmetric arrangement of two boson-condensate
traps with gain and loss of particles \cite{Graefe}.
There is a substantial mathematics literature on the standard $\mathcal{PT}$-symmetric dimer,
concerning stationary points \cite{SP}, periodic orbits \cite{SXK},
conserved quantities \cite{Dimer_integrability},
 the blow-up phenomena \cite{Dima_dimer,Flach} and the geometry of the phase space \cite{Susanto,Flach}.

Equations with parity-time symmetry  
combine properties of dissipative and conservative systems.
A surprising recent result is 
that despite the presence of gain and loss of energy,
a  $\mathcal{PT}$-symmetric system may admit the Hamiltonian formulation \cite{Benfreda,BG}. In particular,
there are Hamiltonian $\mathcal{PT}$-symmetric  dimers; an example 
was produced in \cite{BG}:
\begin{align}
i \frac{du}{dz}    +  v
+  (|u|^2+ 2 |v|^2) u   
+  v^2 u^* &   =                      \phantom{-}   i   \gamma u,     \nonumber 
\\
i \frac{dv}{dz}   +  u 
+  (|v|^2+2|u|^2) v  
+ u^2 v^*  &  =    - i    \gamma v.  
\label{A1}
\end{align}
The  system \eqref{A1}  can be written in the cross-gradient form
\be
i   \frac{du}{dz}=\frac{\partial {\mathcal H}}{\partial {v^*}},\qquad
 i \frac{dv}{dz}= \frac{\partial {\mathcal H}}{\partial {u^*}},
 \label{A5}
\ee
with the Hamilton function
\be
{\mathcal H}= - (|u|^2+|v|^2)(1+u^*v+uv^*)   + i \gamma (uv^*-u^*v).
\label{A2}
\ee

This observation leads one to wonder about the 
 Hamiltonian structure of the most important two-site nonlinear Schr\"odinger equation, namely the standard dimer 
 \eqref{A3}.
 The standard dimer is known to have two functionally independent integrals of motion  \cite{Dimer_integrability,Dima_dimer,Susanto,Flach} ---
yet no explicit Hamiltonian formulation has been put forward so far.

The Hamiltonian structure is a fundamental property of a dynamical system. 
Physically, it establishes regularity of motion (conservation of phase volume
and in some cases, compactness of the accessible part of the phase space) and paves the way for quantisation.
Mathematically, the Hamiltonian structure  implies a deep symmetry of the system which leads to
considerable analytical simplifications and
allows 
to use powerful methods, e.g.   the Hamilton-Jacobi approach
and the Liouville integrability. 

The aim of this note is to establish the Hamiltonian  formulation for the standard dimer.
We also provide the Lagrangian framework for this system.

\section{The Hamiltonian formalism}
The canonical formulation for the standard lossless
dimer 
 is well known
(see e.g. \cite{Eilbeck,Jorgensen,Graefe_thesis}). 
The system  \eqref{A3} with $\gamma=0$ can be written as
\be
i   \frac{du}{dz}= \frac{\partial {\frak H}}{\partial {u^*}},\qquad
 i \frac{dv}{dz}= \frac{\partial {\frak H}}{\partial {v^*}},
 \label{A50}
\ee
where the Hamilton function is
\be
{\frak H}= -(uv^*+u^*v) -  \frac{|u|^4+ |v|^4}{2}.
\label{A200}
\ee
The two pairs of canonical variables are $u$ and $u^*$, $v$ and $v^*$.
We emphasise the difference between Eq.\eqref{A50} and the Hamiltonian structure 
of 
the nonlinearly coupled dimer,  Eq.\eqref{A5}, where the canonical pairs were 
$u$ and $v^*$, $v$ and $u^*$.

Unlike Eq.\eqref{A5}, the structure \eqref{A50} does not survive the addition of the gain-loss terms ---
that is,  $u$ and $u^*$, $v$ and $v^*$ are {\it not\/}  the canonical variables for the system \eqref{A3} with $\gamma \neq 0$.
That being the case, we attempt  to transform both dimers to a new set of coordinates where 
the similarity of the two systems would allow to model
the Hamiltonian structure of the standard $\mathcal{PT}$-symmetric dimer
 on the canonical formulation of
the system \eqref{A1}.  

A natural choice of coordinates is furnished by the 
 Stokes variables,
\be
X=\frac{      u^*v    +   u v^* }{2}, \quad
 Y= i \frac{u^* v-u  v^*}{2}, \quad
   \hfill  Z=\frac{|u|^2-|v|^2}{2}. 
\label{Stokes}
\ee
We also introduce the notation
 \be
 \phi= \phi_1+\phi_2,
 \label{N1} 
 \ee
 where  $\phi_1$ and $\phi_2$ are the arguments of the complex variables $u$ and $v$:
\[
u=  |u| e^{i \phi_1},  \quad
v=|v| e^{i \phi_2}.
\]

  In terms of $X,Y,Z$ and $\phi$, the standard dimer \eqref{A3}  acquires the form
  \begin{subequations} \label{standard}
  \begin{align} 
  {\dot \phi} =   \left( r+ \frac{X}{r} \right) \cosh \psi, 
  \label{Z5} \\
 \label{Z1}
 {\dot X}=- YZ, \quad
 {\dot Y}= (X-1)Z, \quad
 {\dot Z}= \gamma \R  +Y.
 \end{align}
 \end{subequations}
 In Eq.\eqref{Z5},  $r$
is the magnitude of the two-component vector $(X,Y)$:
\be
r= \sqrt{X^2+Y^2},
\label{N200}
\ee 
 whereas in \eqref{Z1}, $\R$ denotes the length of the vector $(X,Y,Z)$:
\[
\R=\sqrt{X^2+Y^2+Z^2}.
\]
 The hyperbolic angle  $\psi$ is defined by 
\be
\R= r \cosh \psi, \quad Z= r \sinh \psi,
\label{N201}
\ee
 and the overdot stands for the derivative with respect to $t=2z$.

 Using the same set of variables, the anharmonically coupled dimer \eqref{A1} becomes
 \begin{subequations} \label{anharmonic}
  \begin{align} 
 {\dot \phi} =
  \left[2 r +  (1+2X)\frac{X}{r}
   \right]
   \cosh \psi ,
 \label{Z2} \\
   {\dot X}=0, \quad
 {\dot Y}= -(1+2X)Z, \\
 {\dot Z}= \gamma \R + (1+2X)Y.
 \end{align}
\end{subequations}

Consider, first, the system \eqref{A1} 
  which admits the Lagrangian and Hamiltonian formulation. The Lagrange
function for  the dimer \eqref{A1} is
\begin{align*}
{\mathcal L} =  \frac{i}{4} \left ( { u}_z v^* -u_z^* v
+ { v}_z u^*  - {v}_z^* u
  \right) \\
+ (1+  uv^*+u^* v) \frac{|u|^2+|v|^2}{2} 
 +  i \frac{\gamma}{2}  (u^* v-uv^*).
\end{align*} 
The corresponding Lagrangian for the system in the form \eqref{anharmonic} results by transforming to $X,Y,Z$, and $\phi$.
Dropping a total derivative, we have
\be
{\mathcal L}  =   {\dot X}
\phi +
{\dot Y} 
\psi 
 + (1+2X)\R  + \gamma Y.
\label{B1}
\ee

We choose $X$ and $Y$ as the coordinates of the  fictitious  classical particle  described by the Lagrangian \eqref{B1}.
The canonical momenta are then
\[
P_X= \frac{\partial {\mathcal L}}{\partial {\dot X}}= \phi,
\quad 
P_Y= \frac{\partial {\mathcal L}}{\partial {\dot Y}}= \psi,
 \]
 and the Hamiltonian of the particle results by the Legendre transform:
\be
\label{Z2}
{\mathcal H}=  - (1+2X)\R - \gamma Y, 
\ee
where
\[
\R=\sqrt{X^2+Y^2} \cosh \psi.
\]
The Hamilton equations
\begin{align*} 
{\dot X} = & \frac{\partial {\mathcal H}}{\partial P_X}, \quad
{\dot P_X}= -  \frac{\partial {\mathcal H}}{\partial X}; \\
{\dot Y} = & \frac{\partial {\mathcal H}}{\partial P_Y},
 \quad
{\dot P_Y}= -  \frac{\partial {\mathcal H}}{\partial Y}
\end{align*} 
reproduce equations \eqref{anharmonic}.

The Hamiltonian \eqref{Z2}  is remindful of the expression for an integral of motion of the standard dimer, $I=-\R-\gamma \theta$,
with the role of $\theta$ being taken over 
by $Y$. Here $\theta$ is one of the two polar coordinates  on the $(X,Y)$-plane  defined by 
\[
X=1 + \rho \sin \theta, \quad 
Y= \rho \cos \theta.
\]
(This choice of polar coordinates is crucial for the elucidation of the geometry of the phase space of the standard dimer
\cite{Flach}.)
 Modelling on the Hamiltonian \eqref{Z2} and noting that 
 \be
 \R= r \cosh \psi, \quad r= \sqrt{\rho^2 +1 + 2 \rho \sin \theta},
 \label{beta}
 \ee
 we define
 \be
  P_\theta= \psi
   \label{Ptheta} 
\ee
and propose the following expression for the Hamiltonian of the standard dimer:
\be \label{Z3}
{ H}=  - \sqrt{\rho^2 +1 + 2 \rho \sin \theta} \cosh P_\theta
 - \gamma \theta.
\ee

The Hamiltonian \eqref{Z3} describes another fictitious            classical          particle ---
 the {\it ``standard"}  particle --- with the coordinates  $\rho$ and $\theta$.
The canonical equations are
\begin{subequations}   \label{HE} 
\begin{align}
{\dot \theta} = \frac{\partial {H}}{\partial P_\theta}=- r \sinh \psi,        \quad
{\dot \rho}= \frac{\partial {H}}{\partial P_\rho}=0,       \label{Z40} \\
{\dot P_\theta} = - \frac{\partial {H}}{\partial \theta}= \gamma + \frac{ \rho \cos \theta}{r} \cosh \psi,  \label{Z70} \\
{\dot P_\rho} = - \frac{\partial {H}}{\partial \rho}=  \frac{\rho + \sin \theta}{ r}  \cosh \psi.   \label{Z60} 
\end{align}
\end{subequations}

The formulas \eqref{Z40}-\eqref{Z70} are equivalent to equations  \eqref{Z1}
while
Eq.\eqref{Z60} can be used to define the momentum $P_\rho$.
Namely, comparing \eqref{Z60} to \eqref{Z5}
and using ${\dot \rho}=0$  yields
\be
P_\rho=- \frac{\phi}{\rho} +\frac{2}{\rho} \int_0^t \R(\tau) d \tau.
\label{Z8}
\ee

\section{The Lagrangian formalism}
The aim of this section is to propose a Lagrangian formulation for the standard 
$\mathcal{PT}$-symmetric dimer. The Lagrangian formulation 
complements the Hamilton equations and offers a number of advantages,
e.g.  the freedom in the coordinate transformations  and access to Noether's  theorem.

Letting
\[
\phi_1-\phi_2= \chi,
\]
the equations of
the standard dimer \eqref{A3} acquire the form
\begin{align}
{\dot r} =- r \sin \chi \sinh \psi,   \nonumber    \\
{\dot \chi}  = (r-\cos \chi)         \sinh \psi,   \nonumber   \\
{\dot \phi} =   (r+\cos \chi)        \cosh \psi,    \nonumber   \\
{\dot \psi} = \gamma + \sin \chi \cosh{\psi}.     
\label{B2}
\end{align}
Here     $\phi$,       $r$, and $\psi$ are defined by \eqref{N1}, \eqref{N200}, and \eqref{N201};
we remind  that $r$ admits a
simple expression in polar coordinates, Eq.\eqref{beta}.

To cast the system \eqref{B2} in the form of the Lagrange-Euler equations for some functional $S=\int L dt$, 
we start with introducing a new variable $\mu$ such that ${\dot \mu}=\R$.
 The constraint ${\dot \mu}-\R=0$ can be incorporated in the system by
means of a  Lagrange multiplier; call it  $\lambda$.
Thus we consider 
the Lagrangian 
\be
L= \frac{2 \mu- \phi}{\rho}   {\dot \rho}
 +         \psi     {\dot \theta}+ \gamma \theta
- \lambda ({\dot \mu} -  \R).
\label{L}
\ee
Here $\rho, \theta, \phi$,  and $\psi$ --- as well as
 $\mu$  and $\lambda$ ---
 are regarded as independent variables,
whereas $\R$ is a function of $\rho$, $\theta$ and $\psi$ given by Eq.\eqref{beta}.

A slightly modified version of \eqref{L} is arrived at by dropping a total derivative:
\be
{\tilde L}= -(2  {\dot \mu}-  {\dot \phi})  \ln \rho
 +         \psi     {\dot \theta}+ \gamma \theta
- \lambda ({\dot \mu} -  \R).
\label{L2}
\ee
The formulation \eqref{L2} makes it obvious that 
the variable $\phi$ is  cyclic; this is a consequence of the U(1) phase invariance of the 
dimer \eqref{A3}. Therefore, $\partial  {\tilde L}/ \partial {\dot \phi}= \ln \rho$ is a conserved quantity:
\begin{subequations}\label{R1}
\be
{\dot \rho}=0.
\ee
The coordinate $\mu$ is also cyclic; hence 
\[
\frac{\partial {\tilde L}}{\partial {\dot \mu}} = -2  \ln \rho  - \lambda
\]
is another integral of motion --- and so is $\lambda$.

The variation with respect to the 
remaining four independent coordinates gives
\begin{align}
  {\dot \theta}=  - \lambda r \sinh \psi, \label{chi} \\
{\dot \phi}  =    2 {\dot \mu}  -          \lambda \frac{ \rho}{r}  (\rho+ \sin \theta) \cosh \psi, \label{rho} \\
{\dot \psi}= \gamma  + \lambda \frac{\rho}{r} \cos \theta \cosh \psi,  \label{theta} 
\end{align}
\end{subequations}
and  ${\dot \mu} =\R$.
The constant $\lambda$   may be chosen arbitrarily;
different choices of $\lambda$ are equivalent up to a rescaling of $t$ and redefinition of $\gamma$.
Choosing $\lambda=1$, one can readily verify that four equations \eqref{R1}
 (with  $\R$  substituted for ${\dot \mu}$) are  equivalent to the system
\eqref{B2}.

\section{Concluding remarks}

In this note, we  have revealed
 the Hamiltonian structure of the standard $\mathcal{PT}$-symmetric 
 dimer, equation \eqref{A3}.
 The Hamilton function is given by \eqref{Z3}; the canonical coordinates are $\rho$ and $\theta$, with the canonical momenta defined by 
 \eqref{Z8} and \eqref{Ptheta}, respectively.

 We have also proposed the Lagrangian formulation for the standard dimer.
 The Lagrange function is in \eqref{L}-\eqref{L2}. 
 Unlike its Hamiltonian formulation, the Lagrangian description 
requires  introduction of an auxiliary
degree of freedom
[accounted for by the variables $\mu$ and $\lambda$  in \eqref{L}].

 We conclude with two remarks.  First, we would like to 
 acknowledge the importance of the Stokes coordinates \eqref{Stokes} 
 that were crucial for our construction.
 Mathematically, the transformation \eqref{Stokes} is an example of 
 the Hopf fibration mapping a 3-sphere (a hypersphere in the four-dimensional space formed by the real
 and imaginary components of $u$ and $v$)
 onto the 2-sphere in the $(X,Y,Z)$-space \cite{Hopf}. 
 In physics, the Hopf map was used to establish the equivalence of two field-theoretic models on the plane,
 the $\mathbf{CP}^1$ model and the $O(3)$  $\sigma$-model \cite{Hopf,CPO}.
 The same transformation is employed in the studies of  quantum  two-level systems  where it was pioneered
 by  Feynman and co-authors \cite{Feynman}. (Accordingly, the $X$, $Y$, and $Z$ are occasionally referred to as the Feynman variables
 \cite{Jorgensen}.) 
 A closely related object is the Bose-Hubbard dimer; in that context, the $X,Y,Z$ triplet is known as the
 Bloch vector  
 \cite{Graefe_thesis}.
  The name {\it Stokes variables\/} hails from optics where
 the $X$,  $Y$,  and $Z$ parameters  are used to  describe the polarization state of electromagnetic radiation. 
 Jensen exploited the Stokes parameters for the analysis of his two-waveguide optical coupler \cite{couplers}.

 Our second remark is on the integrability
 of the standard $\mathcal{PT}$-symmetric dimer.
The fact that a system with two degrees of freedom has two integrals of motion 
is generally insufficient to claim that the system is Liouville integrable. 
Indeed, assigning particular values to the two integrals reduces the motion to a two-dimensional manifold, 
e.g. a genus-two Riemann surface, where the flow may happen  not to be integrable.
However if the system is known to be Hamiltonian, the existence of 
the second integral of motion (which is obviously in involution with the Hamilton function)
implies the complete integrability of the system.
Thus, uncovering the Hamiltonian structure of the standard $\mathcal{PT}$-symmetric dimer
completes  the proof of its integrability that was 
suggested when two conserved quantities were found \cite{Dimer_integrability}.

\acknowledgements
I thank 
 Maxim Pavlov,
Dmitry Pelinovsky, Vyacheslav Priezzhev,  Nikolay Tyurin,
Alexander Vasiliev,  and Alexander Yanovski
for useful conversations. Instructive correspondence with  
Yuri Fedorov and   Boris Malomed is also gratefully acknowledged.
This project was supported by the NRF of South Africa (grants No 85751, 86991, and 87814).




\begin{thebibliography}{99}







\bibitem{couplers}
S Jensen, IEEE Journ Quant Electronics {\bf 18} 1580 (1982);
A. A. Maier, Sov. J. Quantum. Electron. 12, 1490 (1982)








\bibitem{Maier_Snyder} 
A A Maier  Sov. J. Quantum Electron. {\bf 17}   1013   (1987);
Y Chen, A W Snyder, and D N Payne, IEEE Journ Quant Electronics {\bf 28} 239 (1992);
A A Maier,   Phys.-Usp. {\bf 38}  991  (1995)

\bibitem{BEC} 
G. J. Milburn, J. Corney, E. M. Wright, and D. F. Walls, Phys. Rev. A {\bf  55}
4318 (1997);
A. Smerzi, S. Fantoni, S. Giovanazzi, and S. R. Shenoy, Phys. Rev. Lett. 79,
4950 (1997);
S. Raghavan,  A. Smerzi,  S. Fantoni, and S. R. Shenoy,
 Phys Rev A {\bf 59} 620 (1999);
  D. Ananikian and T. Bergeman,
 Phys Rev A {\bf 73} 013604 (2006)
 

 \bibitem{Theocharis}
E. A. Ostrovskaya, Y. S. Kivshar, M. Lisak, B. Hall,   F. Cattani,
and D. Anderson, Phys Rev A {\bf 61} 031601(R) (2000);
G. Theocharis, P. G. Kevrekidis,
D. J. Frantzeskakis, and P. Schmelcher,
Phys. Rev. E {\bf 74}  056608 (2006) 

 \bibitem{Eilbeck}
J C Eilbeck, P S Lomdahl, and A C Scott, Physica D {\bf 16} 318 (1985)

\bibitem{DST}
V. M. Kenkre and D. K. Campbell,  Phys Rev B  {\bf 34} 4959  (1986);
A. C. Scott, J. C. Eilbeck and H. Gilhoj,  Physica D  {\bf 78} 194  (1994)



\bibitem{Tsironis_review} 
D. Hennig, G.P. Tsironis, Phys Rep  {\bf 307}      333  (1999) 


\bibitem{PT}
C M Bender and S Boettcher, Phys Rev Lett {\bf 80} 5243 (1998);
C M Bender,  Contemp. Phys. {\bf 46}   277   (2005); 
 Rep. Prog. Phys. {\bf 70} 947  (2007);
A Mostafazadeh,
Int. J. Geom. Methods Mod. Phys. {\bf 7}  1191 (2010)





\bibitem{PT_couplers}
M Kulishov, J M Laniel, N B\'elanger, J  Aza\~na, D V Plant, Opt Express {\bf 13} 3068 (2005);  
R. El-Ganainy,  K. G. Makris,   D. N. Christodoulides,  and Z. H. Musslimani, Opt Lett {\bf 32} 2632  (2007) 

 \bibitem{Kip}
 C. E. Ruter, K. G. 
Makris, R. El-Ganainy, D. N. Christodoulides, M. Segev, and D. Kip, Nat. Phys. {\bf 6}, 192 (2010)





\bibitem{Dimer_integrability}
H Ramezani, T Kottos, E El-Ganainy, and D N Christodoulides, Phys. Rev. A {\bf 82} 043803 (2010)


\bibitem{SXK} A. A. Sukhorukov, Z. Xu, and Yu. S. Kivshar, Phys. Rev. A {\bf 82} 043818 (2010)





\bibitem{lattices} 
Dmitriev S V, Sukhorukov A A and Kivshar Y S,  Opt. Lett. {\bf 35}   2976  (2010);
S. V. Suchkov, B. A. Malomed,   S. V. Dmitriev, and Yu. S. Kivshar, Phys. Rev.  E {\bf 84} 046609 (2011);
 R Driben and B A Malomed, Opt. Lett. {\bf 36} 4323 (2011);
N. V. Alexeeva, I. V. Barashenkov,  A. A. Sukhorukov, and Yu. S. Kivshar,  Phys. Rev. A {\bf 85} 063837 (2012);
I. V. Barashenkov,  S. V. Suchkov,   A. A. Sukhorukov,  S. V. Dmitriev, and Yu. S. Kivshar, Phys. Rev. A {\bf 86}  053809 (2012);
Zezyulin D A and Konotop V V,    Phys. Rev. Lett. {\bf 108} 213906 (2012);
 R L Horne, J Cuevas, P G Kevrekidis, N Whitaker,
F Kh Abdullaev,   and D J Frantzeskakis, 
Journ Phys A: Math Theor  {\bf 46}    485101      (2013);
A. J. Mart\'inez, M. I. Molina, S. K. Turitsyn, and Y. S. Kivshar,
 arXiv:1405.3032 [physics.optics] (2014)

 
\bibitem{Graefe}
E. M. Graefe,
H. J. Korsch, and A. E. Niederle, 
Phys. Rev. Lett. {\bf 101} 150408 (2008);
E. M. Graefe,
H. J. Korsch, and A. E. Niederle, 
 Phys. Rev. A {\bf  82}, 013629 (2010);
 E.-M. Graefe, J. Phys. A: Math. Theor. {\bf 45}  (2012) 444015





\bibitem{SP} 
K. Li and P. G. Kevrekidis, Phys. Rev. E {\bf 83} 066608 (2011);
A. S. Rondrigues, K. Li, V. Achilleos, P. G. Kevrekidis, D. J. Frantzeskakis, C. M. Bender,
Romanian Reports in Physics {\bf 65} 5 (2013);
J Cuevas,  P G Kevrekidis,   A Saxena,  A Khare,  Phys Rev A {\bf 88} 032108 (2013);
M. Duanmu,    K. Li,   R. L. Horne,  P. G. Kevrekidis, 
and N. Whitaker, Phil Trans R
Soc A { \bf 371}  20120171 (20130 



\bibitem{Dima_dimer}
P G Kevrekidis, D E Pelinovsky, and D Y Tyugin, J. Phys. A {\bf 46} 356201 (2013)


\bibitem{Flach}
I. V. Barashenkov,    G. S. Jackson,  and S. Flach,
Phys Rev A  {\bf 88}     053817 (2013)

 
\bibitem{Susanto}
J Pickton and H Susanto, Phys. Rev. A {\bf 88}  063840 (2013)


\bibitem{Benfreda} C. M. Bender,   M. Gianfreda, \c{S}. K. \"Ozdemir, B. Peng, and L. Yang,
 Phys. Rev. A {\bf 88}, 062111 (2013)





\bibitem{BG} 
I V Barashenkov and M Gianfreda, J Phys A: Math Theor {\bf 47}                    282001        (2014)






\bibitem{Jorgensen}
M F J{\o}rgensen, P L Christiansen and I Abou-Hayt, Physica D {\bf 68}    180  (1993);
M F J{\o}rgensen and  P L Christiansen, Chaos, Solitons and Fractals  {\bf 4} 217 (1994);
M F J{\o}rgensen and  P L Christiansen, J. Phys. A: Math. Gen. {\bf 31}  969 (1998) 







\bibitem{Graefe_thesis} 
E.-M. Graefe, 
Quantum-classical correspondence for a Bose-Hubbard dimer and its nonhermitian generalisation.
 PhD thesis,
  Technischen Universit\"at Kaiserslautern (2009);
aleph.physik.uni-kl.de/$\sim$korsch/diss/diss$\_$graefe.pdf

\bibitem{Hopf}
M. Bergeron, G. W. Semenoff, and R. R. Douglas, Int. J. Mod. Phys. A {\bf 7} 2417 (1992)


\bibitem{CPO}
R Rajaraman, Solitons and Instantons. An Introduction to Solitons and Instantons in Quantum Field Theory.
North-Holland, 1982
(section 4.5);  
W J Zakrzewski, Low Dimensional Sigma Models. Adam Hilger, 1989 (section 3.2);
M Shifman, Advanced Topics in Quantum Field Theory. Cambridge University Press, 2012
(section 27.2)



\bibitem{Feynman} 
R. P. Feynman, F.  L. Vernon Jr.,  and R. W. Hellwarth,  J. Appl. Phys. {\bf 28}   49 (1957)





\end{thebibliography}
\end{document}